\documentclass{ecai} 
\usepackage{latexsym}
\usepackage{amssymb}
\usepackage{amsmath}
\usepackage{amsthm}
\usepackage{booktabs}
\usepackage{enumitem}
\usepackage{graphicx}
\usepackage{color}
\usepackage{multirow}
\usepackage{graphicx}
\usepackage{graphics}      % for EPS, load graphicx instead 
\usepackage{color}
\usepackage{booktabs}
\usepackage{float}
\usepackage[export]{adjustbox}
\usepackage{soul}

\usepackage{caption,subcaption,graphicx}
\newcommand{\BibTeX}{B\kern-.05em{\sc i\kern-.025em b}\kern-.08em\TeX}

\begin{document}

%%%%%%%%%%%%%%%%%%%%%%%%%%%%%%%%%%%%%%%%%%%%%%%%%%%%%%%%%%%%%%%%%%%%%%%%

\begin{frontmatter}

%%% Use this command to specify your submission number.
%%% In doubleblind mode, it will be printed on the first page.

\paperid{123} 

%%% Use this command to specify the title of your paper.

\title{Interactive Example-based Explanations to Improve Health Professionals' Onboarding with AI for Human-AI Collaborative Decision Making}

%%% Use this combinations of commands to specify all authors of your 
%%% paper. Use \fnms{} and \snm{} to indicate everyone's first names 
%%% and surname. This will help the publisher with indexing the 
%%% proceedings. Please use a reasonable approximation in case your 
%%% name does not neatly split into "first names" and "surname".
%%% Specifying your ORCID digital identifier is optional. 
%%% Use the \thanks{} command to indicate one or more corresponding 
%%% authors and their email address(es). If so desired, you can specify
%%% author contributions using the \footnote{} command.

%\author[A]{\fnms{Min Hun}~\snm{Lee}\orcid{....-....-....-....}\thanks{Corresponding Author. Email: mhlee@smu.edu.sg}\footnote{Equal contribution.}}
%\author[A]{\fnms{Min Hun}~\snm{Lee}\orcid{....-....-....-....}\thanks{Corresponding Author. Email: mhlee@smu.edu.sg}}
%\author[B]{\fnms{Renee Bao Xuan}~\snm{Ng}\orcid{....-....-....-....}\footnotemark}
%\author[A]{\fnms{Renee Bao Xuan}~\snm{Ng}\orcid{....-....-....-....}}
%\author[C]{\fnms{Shamala D/O}~\snm{Thilarajah}\orcid{....-....-....-....}} 
%\author[D]{\fnms{Silvana Xinyi}~\snm{Choo}\orcid{....-....-....-....}} 
\author[A]{\fnms{Min Hun}~\snm{Lee}\thanks{Corresponding Author. Email: mhlee@smu.edu.sg}}
\author[A]{\fnms{Renee Bao Xuan}~\snm{Ng}}
\author[B]{\fnms{Silvana Xinyi}~\snm{Choo}} 
\author[B]{\fnms{Shamala}~\snm{Thilarajah}} 

\address[A]{Singapore Management University}
\address[B]{Singapore General Hospital}

%%% Use this environment to include an abstract of your paper.

\begin{abstract}
A growing research explores the usage of AI explanations on user's decision phases for human-AI collaborative decision-making. However, previous studies found the issues of overreliance on `wrong' AI outputs. In this paper, we propose interactive example-based explanations to improve health professionals' onboarding with AI for their better reliance on AI during AI-assisted decision-making. We implemented an AI-based decision support system that utilizes a neural network to assess the quality of post-stroke survivors' exercises and interactive example-based explanations that systematically surface the nearest neighborhoods of a test/task sample from the training set of the AI model to assist users' onboarding with the AI model. To investigate the effect of interactive example-based explanations, we conducted a study with domain experts, health professionals to evaluate their performance and reliance on AI. 
Our interactive example-based explanations during onboarding assisted health professionals in having a better reliance on AI and making a higher ratio of making `right' decisions and a lower ratio of `wrong' decisions than providing only feature-based explanations during the decision-support phase. Our study discusses new challenges of assisting user's onboarding with AI for human-AI collaborative decision-making. 
\end{abstract}

\end{frontmatter}

%%%%%%%%%%%%%%%%%%%%%%%%%%%%%%%%%%%%%%%%%%%%%%%%%%%%%%%%%%%%%%%%%%%%%%%%

\section{Introduction}
Advanced artificial intelligence (AI) has been increasingly being explored and considered to provide data-driven insights for improving various decision-making tasks (e.g. social services \cite{zavrvsnik2020criminal} and health \cite{beede2020human,cai2019human,lee2021human}). Researchers have investigated how to effectively form human-AI teams \cite{cai2019human,topol2019high,lee2021human} instead of applying a fully autonomous approach for AI systems in high-stake contexts. However, it remains challenging to explain the rationale of an AI output \cite{london2019artificial,rajpurkar2022ai,cai2019human}, build an appropriate trust on AI \cite{bansal2021does,wang2021explanations,bussone2015role,buccinca2021trust}, and integrate these AI systems in practice \cite{sutton2020overview,khairat2018reasons,kohli2018cad}.

To address these challenges of forming effective human-AI teams, a growing body of research has explored various explainable AI techniques \cite{lipton2018mythos,preece2018asking,wang2019designing,lakkaraju2020explaining} and the interactive visualization techniques of high-dimensional embedding data \cite{boggust2022embedding,liu2016visualizing,abdi2010principal,becht2019dimensionality,van2008visualizing}. In addition, AI and HCI researchers have conducted studies \cite{liao2021human,cai2019hello,cai2019human,beede2020human,lee2021human,wang2019designing,holstein2023toward} with the stakeholders to understand what they need and explore how they can effectively interact with AI explanations in specific applications. Some studies have shown the utility of AI explanations and interactive visualizations \cite{cai2019human,lai2019human,wang2021explanations} for human-AI teams with improved accuracy. However, previous studies also discussed an issue and harmful effect of using AI explanations to include user's overreliance on AI even when it is wrong \cite{bussone2015role,bansal2021does,buccinca2021trust,lee2023counter}. In addition, most studies have focused on utilizing AI explanations at decision support phases
\cite{lai2019human,bansal2021does,lai2019human,buccinca2021trust,wang2021explanations,lee2021human}. There has been limited understanding of how AI explanations can be used to assist user's onboarding phases when an AI system has been first introduced to a user \cite{cai2019hello}. 

In this work, we focused on the context of a clinical decision-making task (i.e. physical stroke rehabilitation assessment) and investigated how interactive example-based can be used to support user's onboarding with AI for human-AI collaborative decision-making. 
To ground this research, we first interviewed domain experts (i.e. health professionals) to rank and discuss three widely used AI explanations to support their onboarding with AI. Building upon the findings from the interviews and the previous research of human-AI interaction \cite{GooglePAIR2019,amershi2019guidelines,cai2019hello}, we created an AI-based decision support system that leverages a neural network to assess the quality of post-stroke survivor's exercises and provides interactive example-based explanations to assist user's onboarding with AI and feature-based explanations to support a user's decision-making task. 

Given a new patient's exercise motion, the system provides interactive example-based explanations (Figure \ref{fig:system_example}) that identify the nearest $k$-neighbourhoods of the new data from the training set of the AI model and visualize their embedding spaces along with AI outputs and ground truth labels. As interactive example-based explanations allow users to review how data has been represented and how an AI model performed on the nearest neighborhoods, the interactive example-based explanations can potentially assist users' onboarding with AI for their better reliance on AI during their AI-assisted decision-making. 

To evaluate the effect of interactive example-based explanations, we conducted an experiment with sixteen domain experts, health professionals (i.e. therapists) and analyzed their reliance on AI during their human-AI collaborative decision-making. The results showed that user's onboarding phases with our interactive example-based explanations assisted therapists to have a better-calibrated reliance on AI and making a higher ratio of making `right' decisions and a lower ratio of `wrong' decisions than providing only feature-based explanations during the decision-support phase.
Our findings suggest the potential of interactive example-based explanations to support user's onboarding with AI and point to challenges on how AI explanations can be utilized for onboarding users with an AI model and more effective human-AI collaborative decision-making in high-stake domains (e.g. health) \cite{cai2019hello,lee2021human,cai2019human,beede2020human}.

\section{Related Work}
We describe prior work at the intersection of explainable AI and visualization techniques and empirical studies of human-AI collaborative decision-making. 

\subsection{Explainable AI \& Visualization Techniques}
Researchers have explored human-AI collaborative decision-making, in which the AI system provides users data-driven insights on a decision-making task to assist and improve human's final decision-making \cite{lai2021towards,cai2019hello,lee2021human,zavrvsnik2020criminal,beede2020human}. However, users have difficulty with understanding why the AI system with a complex algorithm provides a certain outcome \cite{london2019artificial,rajpurkar2022ai}. They may resist and abandon the usage of these systems in practice \cite{khairat2018reasons}. To this end, researchers have explored diverse explainable AI (XAI) and visualization techniques to improve user's understanding of how AI/ML-based models reach their outcomes \cite{lipton2018mythos,lakkaraju2020explaining,boggust2022embedding,lee2024towards}. 

XAI techniques can be categorized into inherently interpretable models, such as linear regression models, rule-based models, and decision trees, and post-hoc methods that generate an approximate of the model's decision logic by producing understandable representation, such as relevant examples or feature importance scores \cite{lakkaraju2020explaining}. 
In this work, we focus on exploring two widely used post-hoc XAI techniques: an important feature explanation and an example-based explanation. Important feature explanations compute the contributions of input features to a model output and present the list of identified features or highlighted pixels \cite{lakkaraju2020explaining,gilpin2018explaining,ribeiro2016should}. Example-based explanations identify and present samples that are the most relevant and influential to an AI output \cite{lakkaraju2020explaining,gilpin2018explaining,cai2019effects}.

In addition, various techniques have been proposed to visualize and interpret high-dimensional input or representations of AI/ML models \cite{liu2016visualizing}. Specifically, researchers have projected the high-dimensional input data or embedding representations of AI/ML models into two or three reduced dimensions using dimensional reduction techniques (e.g. principal component analysis) \cite{abdi2010principal,becht2019dimensionality,van2008visualizing} to visualize reduced dimensions on an interactive tool \cite{boggust2022embedding}.

\subsection{Studies of Human-AI Collaborative Decision Making}
For human-AI collaborative decision-making, researchers have suggested engaging with the stakeholders, understanding what users need, and exploring how they can effectively interact with AI explanations in specific applications \cite{cai2019hello,cai2019human,beede2020human,lee2021human} 
In addition, there have been increasing studies to explore the effect of AI explanations for various decision-making tasks (e.g. deception detection \cite{lai2019human}, cancer diagnosis \cite{cai2019human}, and stroke rehabilitation assessment \cite{lee2021human}).
Some of the previous studies discussed that providing explanations could lead to a harmful effect of user's over-reliance on the system \cite{bussone2015role,bansal2021does,lee2023counter,bansal2021does}.

For the issue of overreliance on AI, researchers have conducted various empirical studies to understand the effect of strategies or factors (e.g. cognitive forcing intervention \cite{buccinca2021trust} or presenting AI explanations at the decision-making phase \cite{lee2023counter,vasconcelos2023explanations}. For instance, Buccinca et al. \cite{buccinca2021trust} discussed the cognitive forcing intervention, such as slowing down the process and asking the person to make a decision before seeing the AI recommendation. Lee and Chew \cite{lee2023counter} demonstrated the potential of counterfactual explanations to increase users' analytical reviews on AI outputs and reduce their overreliance on AI during human-AI collaborative decision-making. 
However, a growing research work has mostly studied the issue of AI overreliance by presenting AI explanations at the decision support phase \cite{lai2019human,bansal2021does,lai2019human,buccinca2021trust,wang2021explanations}. We have a limited understanding of how we can effectively support a user's onboarding phase with AI \cite{cai2019hello} when an AI system has been first introduced to a user for the user's trust and reliance on AI. 

In this work, we focused on the context of the AI-assisted clinical decision-making task (i.e. physical stroke rehabilitation assessment) and contributed to exploring the effect of an interactive AI explanation to support user's onboarding with AI for human-AI collaborative decision-making. 
To this end, we engaged with the domain experts to seek their opinions on how AI explanations can be used to support user's onboarding. In addition, compared to other works that utilize a mock-up decision support system that operates with the wizard-of-oz approach \cite{bussone2015role} or a simulated AI model \cite{buccinca2021trust,vasconcelos2023explanations}, this work utilized the dataset of 15 post-stroke survivors to implement AI model outputs and explanations and conducted an experiment with domain experts.

\section{Study Designs}
In this work, we explore the effect of interactive AI explanations to support users' onboarding with AI for human-AI collaborative decision-making. Building upon increasing research on explainable AI techniques for AI-assisted decision making \cite{bussone2015role,buccinca2021trust,wang2021explanations,cai2019human,lee2021human} and previous research work that describes the importance of communicating the strength of AI for onboarding with AI \cite{cai2019hello}, we focused on studying how an explainable AI technique can be used to introduce the strength of AI for user's onboarding with AI and understanding its effect during human-AI collaborative decision making (i.e. physical stroke rehabilitation assessment). 

To investigate this research question, we first conducted semi-structured interviews with domain experts (i.e. therapists) to probe their opinions on how an explainable AI technique can be used to communicate the strengths and limitations of AI for onboarding. Building upon the findings from the interviews, we created an interactive example-based explanation to support user's onboarding with an AI-based system. We then experimented with therapists to examine the effect of an interactive example-based explanation for their trust and reliance on AI during decision-making tasks of assessing post-stroke survivors' quality of motion. The study materials and procedures were approved by the Institutional Review Board (IRB).

\subsection{Study Context}
This work focuses on the context of a clinical decision-making task of assessing the quality of motion of post-stroke survivors. We built upon the previous research on AI-assisted decision-making on stroke rehabilitation assessment \cite{lee2020co} to specify an upper limb exercise and performance components of rehabilitation assessment. For an exercise, a post-stroke survivor has to raise the survivor's wrist to the mouth as if drinking water. The performance components of rehabilitation assessment include `Range of Motion (ROM)' that checks how closely a post-stroke survivor achieves the target position of an exercise and `Compensation' that checks whether a post-stroke survivor involves any unnecessary, compensatory joint movements to perform an exercise (e.g. leaning to the side) \cite{lee2020co}.

\subsection{Interviews about Onboarding with AI}
To understand how an explainable AI technique can be used as an onboarding material to communicate the strengths and limitations of AI, we conducted semi-structured interviews with ten therapists, who have experience with managing stroke rehabilitation (Appendix. Table \ref{tab:demographics_detailed_interview}) \cite{lee2024improving}. We recruited the participants through advertisements sent to the hospital staff, the mailing lists, and the contacts of the research team. 

For the interviews, we first introduced an AI-based system and three AI explanations for physical stroke rehabilitation assessment and asked the participants to rank which AI explanations are useful to support onboarding (i.e. when a user first reviews and interacts with AI to understand the strength of AI) and decision support (i.e. when a user review AI outputs for a decision-making task) \cite{lee2024improving}. All interview sessions were conducted remotely on a video platform for 60 to 80 minutes.

To introduce an AI-based system AI explanations, we utilized existing guidelines for human-AI interaction \cite{amershi2019guidelines,GooglePAIR2019,cai2019hello}, an AI model card \cite{mitchell2019model}, and tutorials of AI explanations \cite{lakkaraju2020explaining} to create introduction materials of an AI-based decision support system for physical stroke rehabilitation assessment. Also, we had discussions with domain experts in stroke rehabilitation to refine our introduction materials before conducting the interviews. 

For onboarding, therapists \textbf{ranked an example-based explanation (36.7\%) as the most useful}, a counterfactual explanation (35.0\%) as the second most useful, and a feature importance explanation (28.3\%) as the third most useful \cite{lee2024improving}. For decision support, therapists ranked an example-based explanation and a feature importance explanation as equally useful (33.9\%) and a counterfactual explanation (32.2\%) as the third most useful \cite{lee2024improving}. For onboarding, therapists described that they want to \textit{``briefly validate the correctness of AI outputs to develop a trust with AI''}. An example-based explanation is useful for onboarding as \textit{``reviewing a pool of similar samples''} is \textit{``easier to interpret than others''} \cite{lee2024improving}. In addition, therapists suggested presenting benchmarkable information to understand the strength of an AI model, such as characterizing the conditions of similar post-stroke survivors, how much therapists had agreed on assessment, and how well the AI model can replicate therapist's assessment scores \cite{lee2024improving}. 

\begin{figure*}[htp]
\centering 
\begin{subfigure}[t]{0.5\textwidth}
\centering
  \includegraphics[width=1.0\columnwidth]{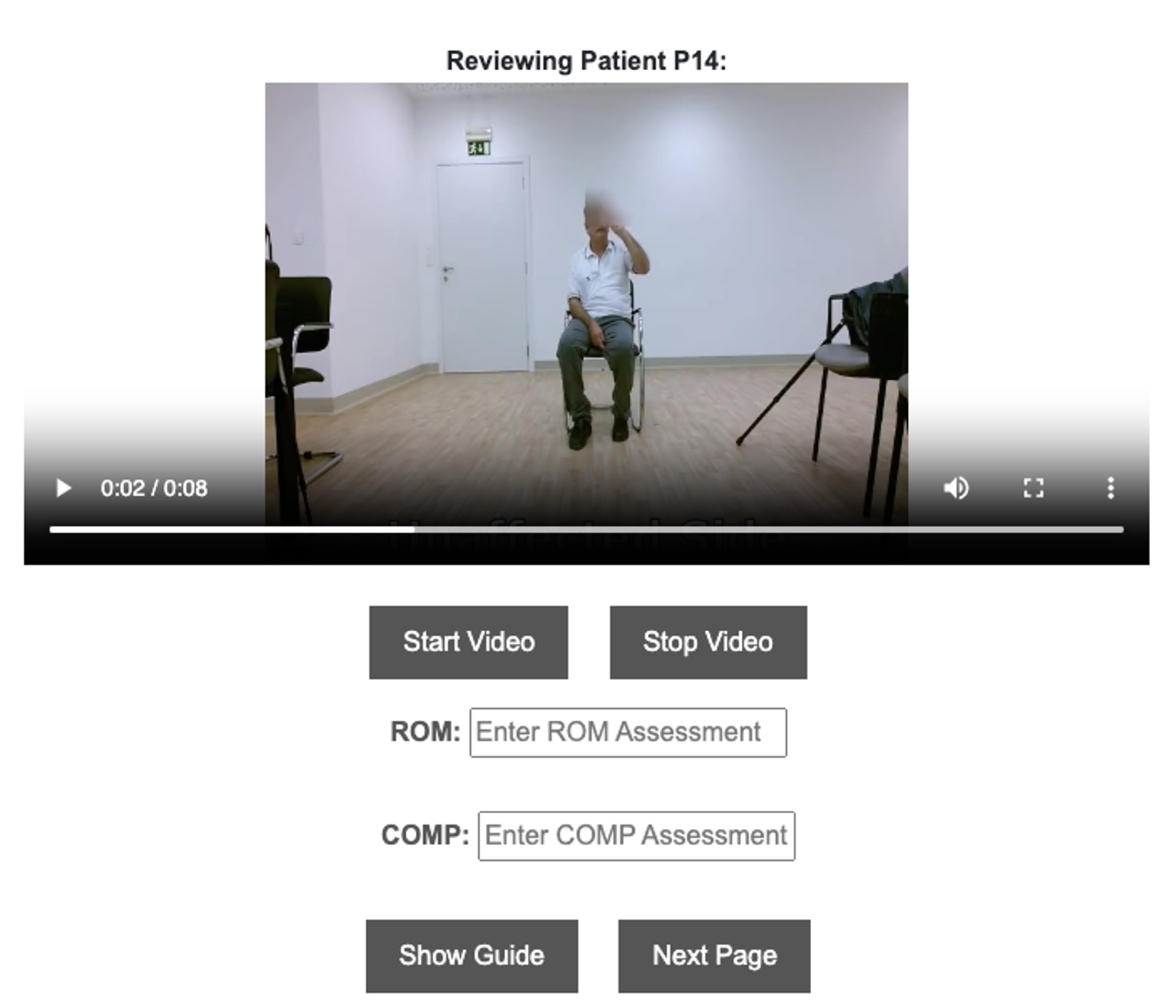}
  \caption{}
  \label{fig:system_video}
\end{subfigure}\vspace{5mm}
\begin{subfigure}[t]{0.55\textwidth}
  \centering
  \includegraphics[width=1.0\columnwidth]{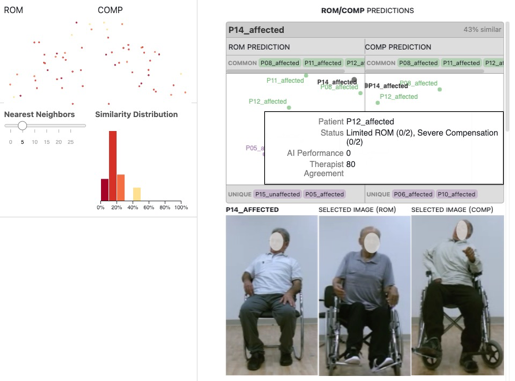}
  \caption{}
  \label{fig:system_example}
\end{subfigure}\vspace{5mm}
\begin{subfigure}[t]{0.4\textwidth}
  \centering
  \includegraphics[width=1.0\columnwidth]{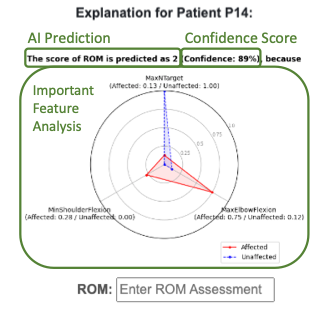}
  \caption{}
  \label{fig:system_feature}
\end{subfigure}\vspace{5mm}
\caption{The AI-based decision support system for physical stroke rehabilitation assessment. This system presents (a) the video of a post-stroke survivor's exercise, (b) interactive example-based explanations along with embedding visualization to assist user's onboarding with AI, and (c) an important feature explanation that compares a post-stroke survivor's unaffected and affected sides using the top three most important features to assist rehabilitation assessment}\label{fig:system}
\end{figure*}
\section{System Implementations}
Informed by the findings of the interviews with domain experts, we created an AI-based decision support system (Figure \ref{fig:system}). Given a video of a post-stroke survivor's rehabilitation exercise, this system utilizes a neural network model to classify the quality of motion. In addition, the system includes an interactive example-based explanation (Figure \ref{fig:system_example}) for facilitating the user's onboarding with the AI system and an important feature explanation (Figure \ref{fig:system_feature}) for assisting the user's decision-making on rehabilitation assessment. In the following section, we described the dataset of our study and the implementations of an AI model, AI explanations, and the interface in detail. 

\subsection{Dataset}
This work utilizes the dataset of a \textit{``Bring a cup to the mouth''} upper-limb exercise from 15 post-stroke survivors with diverse status of functional abilities \cite{lee2019learning}. The dataset contains (1) 300 videos of 15 post-stroke survivors, who performed ten trials of the exercise using their unaffected and affected side by stroke, (2) estimated joint positions of their exercise motions using a Kinect sensor v2, and (3) the annotations by the expert therapist, who utilized clinically validated assessment tool \cite{gladstone2002fugl} to check the status of post-stroke survivors and another therapist, who had not had any interactions with 15 post-stroke survivors. For the annotations on performance components of rehabilitation assessment, therapists individually watched the videos of post-stroke survivors without reviewing any AI outputs.

\subsection{AI Model}
Following the previous research on rehabilitation assessment \cite{lee2019learning}, we processed the estimated joint positions of post-stroke survivors' exercises to extract various kinematic features. The kinematic features of the `Range of Motion' (ROM) include joint angles, such as elbow flexion, shoulder flexion, and elbow extension, and normalized relative trajectory (i.e. the Euclidean distance between two joints - head and wrist; head and elbow), and the normalized trajectory distance (i.e. the absolute distance between two joints - head and wrist, shoulder and wrist) in the x, y, and z coordinates \cite{lee2019learning}. The features of the `Compensation' include the normalized trajectories, which indicate the distances between joint positions of the head, spine, and shoulder in the x, y, and z coordinates from the initial to the current frame over the entire exercise motion \cite{lee2019learning}.

Given the extracted kinematic features and labels of post-stroke survivors' exercises, we implemented a feed-forward neural network (NN) model to classify the quality of post-stroke survivors' motion using Pytorch libraries \cite{paszke2019pytorch}, following its outperformance shown in the previous research \cite{lee2019learning}. For the labels, we utilized the labels by the expert therapist, who conducted the clinically validated assessment test.
We grid-searched various architectures (i.e. one to three layers with 32, 64, 128, 256, and 512 hidden units) and different learning rates (i.e. 0.0001, 0.0005, 0.0001, 0.005, 0.001) while training a feed-forward NN model with cross-entropy loss and the mini-batch size of 1 and epoch of 4. 

For training and evaluating the model, we utilized the leave-one-subject-out cross-validation, where we trained the model with data from all post-stroke survivors except one post-stroke survivor and tested the model with data from the held-out post-stroke survivor. The final model architectures and learning rates are three layers with 256 hidden units and 0.005 of the learning rate for the ROM and three layers with 64 hidden units and 0.005 of the learning rate. The models achieved 82\% F1-score and 77\% F1-score to replicate therapists' assessment on `ROM' and `Compensation' components respectively.

\subsection{AI Explanations and Interface}
Our interactive system interface presents an AI prediction on performance components of rehabilitation assessment along with its confidence score \cite{amershi2019guidelines}. In addition, our system provides 1) interactive example-based explanations for improving users' onboarding with AI and 2) feature-based explanations for assisting their decision-making. For the implementation of our system interface, we utilized the python, flask \cite{grinberg2018flask}, HTML, and javascript libraries. 

\subsubsection{Interactive Example-based Explanation for Onboarding}
Our interactive example-based explanation (Figure \ref{fig:system_example}) aims to assist users in onboarding with AI and developing initial trust with AI by reviewing similar cases of a test/query input. Specifically, the interactive example-based explanation shows the global views of the embedding spaces of the entire data \cite{boggust2022embedding} and the local views of the embedding spaces of $k$-nearest neighbors \cite{boggust2022embedding} of a case to be reviewed. In addition, our interactive example-based explanation shows common (green color) and unique (purple color) neighbor lists of embedding data from two performance components (`Range of Motion' - ROM and `Compensation' - COMP). 

A user can specify the embedding space to review and the number of $k$-nearest neighbors. A user can click embedding data to review the image of a neighbor post-stroke survivor. A user can hover around embedding data to quickly review the benchmarkable information of a neighbor on a tooltip (Figure \ref{fig:system_example}). The benchmarkable information includes the status of a neighboring post-stroke survivor, the performance of the AI model, and the therapists' agreement level on selected neighboring post-stroke survivor's data.

We hypothesize that by reviewing AI performance and therapists' agreement on nearest neighbor data, a user can have a better understanding of the strength of AI and build a better calibrated initial trust in AI for effective human-AI collaborative decision-making.

For an example-based explanation, we explored the representation of input and intermediate layers of the feed-forward NN models and explored principal component analysis (PCA) \cite{abdi2010principal}, Uniform Manifold Approximation and Projection (UMAP) \cite{becht2019dimensionality}, and t-distributed Stochastic Neighbor Embedding (t-SNE) \cite{van2008visualizing} to compute embedding data while implementing a $K$-nearest neighbor classifier over various sizes of $k$ (i.e. 5, 10, 15, 20, 25, 30) with cosine or Euclidean distance \cite{turney2010frequency}. Based on the experimental results, we utilized the default value of $k$ as 5, the Euclidean distance metric, UMAP, and the first input layers of the feed-forward Neural network models as embedding data. 

For presenting embedding data, we utilized the Embedding Comparator \cite{boggust2022embedding} and revised it to present images of similar data points when a user clicks a data point and show benchmarkable information on a tooltip when a user hovers a data point.

\subsubsection{Feature-based Explanation for Decision-Support}
Among various types of AI explanations, we built upon the previous research that describes therapists' preferences in reviewing feature-based explanations on rehabilitation assessment tasks \cite{lee2020co} and the findings of our interviews and utilized a feature-based explanation to assist users' decision-making tasks. 

For a feature-based explanation (Figure \ref{fig:system_feature}), we first identify user-specific important features of rehabilitation assessment using the trained feed-forward neural network models and the SHAP library \cite{NIPS2017Shap}. We then utilized only the top three features to avoid overwhelming users with a list of features \cite{amershi2019guidelines}. Following the practices of therapists \cite{lee2020co}, we utilized a radar chart to compare these features on post-stroke survivors' unaffected and affected sides by stroke to assist users' decision-making tasks.

\section{Experiments}
Given the issue of overreliance on AI \cite{wang2021explanations,lai2021towards,lee2023counter}, this work hypothesizes that our interactive example-based explanation will enable users to understand the strengths and limitations of an AI model and develop a better-calibrated trust and reliance on AI for human-AI collaborative decision making. 

We specified two conditions (``Features'' without interactive example-based explanation and ``Examples + Features'') and conducted a with-in subject study with sixteen health professionals to explore the effectiveness of our interactive example-based explanations on users' AI-assisted decision-making. 

\begin{itemize}
    \item \textbf{``Features''}: the baseline, controlled condition of an AI-based decision support system that presents videos of post-stroke survivors' exercises (Figure \ref{fig:system_video}), AI predicted assessment scores, and an important feature explanation (Figure \ref{fig:system_feature}) without interactive example-based explanations
    \item \textbf{``Examples + Features''}: the experimental condition of an AI-based decision support system that includes the same functionalities of the baseline condition along with an additional, interactive example-based explanation for onboarding with the AI model (Figure \ref{fig:system_example})
\end{itemize}

As previous work describes therapists' preferences to review feature-based explanations and find evidence \cite{lee2020co}, we included the feature-based explanations by default to confirm therapists' hypothetical assessment with an AI explanation \cite{wang2019designing}. In addition, an interactive example-based explanation is included in the experimental condition to support user's onboarding with AI. 
In our study, we referred to two systems as ``Condition A'' and ``Condition B'' to avoid biasing participants. We referred to these conditions as ``Features'' and ``Examples + Features'' for clarity throughout the paper.

\begin{comment}
In addition, we specified two groups of participants to understand the effect of tutorials on AI for user's reliance on AI during human-AI collaborative decision-making.
\begin{itemize}
    \item \textbf{``NTut''}: the baseline, controlled group of participants, who have been introduced to the AI model (i.e. inputs and outputs) and AI explanations without an AI model card \cite{mitchell2019model} that describes the performance of the AI model
    \item \textbf{``YTut''}: the experimental group of participants, who have been introduced to the AI model and AI explanation similar to the baseline group along with the AI model card that describes the performance of the AI model on the dataset (Figure \ref{fig:tutorial_performance})
\end{itemize}
\end{comment}

For the study, we recruited 16 health professionals, therapists, who have experience with stroke rehabilitation through an advertisement sent to the hospital staff, the mailing list, and the contacts of the research team.
We described the detailed demographics in the Appendix. Table \ref{tab:demographic_detailed}.

\subsection{Protocol}

We conducted a within-subject experiment to understand the effect of interactive example-based explanations on users' reliance on AI during human-AI collaborative decision-making. After a participant completed the informed consent form, each participant was randomly assigned to (i) either first use the AI with only important feature explanations (Condition A - `Features') without interactive AI explanations and then AI with interactive example-based explanations for onboarding and important feature explanations (Condition B - `Examples + Features') or vice-versa. 

On each condition, the participants conducted two sub-tasks of completing the rehabilitation assessment on their assigned cases (i) without reviewing AI outputs and explanations (Figure \ref{fig:system_video}) and (ii) after reviewing AI outputs and explanations. In each condition, the participant conducted eight initial assessments and eight final assessments on post-stroke survivors' exercises after reviewing AI outputs and explanations. Among eight assigned cases, we included the cases of 4 `right' AI outputs and 4 `wrong' AI outputs by the trained feed-forward neural network models to investigate the effect of AI explanations on user's overreliance on `wrong' AI outputs.

We counterbalanced the assigned cases of each condition and randomized the order of the two conditions and the presentations of assigned cases of post-stroke survivors. All participants received a fixed compensation for their participation based on the rate recommended by the domain experts. 

\subsection{Evaluation Metrics}
We built upon previous studies on human-AI collaborative decision making \cite{lai2021towards,cai2019human,lee2023counter} and utilized the following evaluation metrics: 1) performance, 2) ratio of `right' and `wrong' decisions, and 3) the duration of decision-making tasks.

\textbf{1) Performance:}
We utilized the annotations of a therapist, who conducted the clinically validated functional assessment test, as ground truth scores and measured the participants' performance on decision-making tasks before and after reviewing AI outputs and explanations \cite{lai2021towards}.

\textbf{2) Ratio of `Right' and `Wrong' Decisions:}
We computed the ratio of `right' and `wrong' decisions by the participants using two variants of the system: "Features" without interactive example-based explanations and "Examples + Features". In addition, we further analyzed the ratio of (1) agreeing with `right' AI outputs, (2) rejecting `wrong' AI outputs, (3) agreeing with `wrong' AI outputs (i.e. overreliance), and (4) rejecting `right' AI outputs (i.e. underreliance). 

\textbf{3) Duration of Decision-Making:}
Our systems computed the estimated duration of decision-making by measuring the time from reviewing a video for the assessment to submit an assessment score.

\section{Results and Discussion}
We present the results of our evaluation metrics (i.e. performance, ratio of `right' and `wrong' decisions, and duration of decision making). We refer to the decision-making of participants without reviewing AI outputs and explanations as \textbf{\textit{``Human''}} and with reviewing AI outputs and explanations as \textbf{\textit{``Human + AI''}}. In addition, we refer to Condition A as \textbf{\textit{``Features''}} where participants interact with the AI system with only important feature explanations, and Condition B as \textbf{\textit{``Examples + Features''}} where participants interact with the interactive AI system with example-based explanations, the visualizations of embeddings, and important feature explanations. 

\subsection{Performance}
Table \ref{tab:results_performance} summarizes the average performance (i.e. F1-score) of rehabilitation assessment tasks without AI outputs (`Human') and with AI outputs (`Human + AI') by participants. 

\begin{table}[h]
\centering
\caption{Performance of rehabilitation assessment tasks using AI with only Features (Condition A) and AI with Examples + Features (Condition B)}
\label{tab:results_performance}
\resizebox{\columnwidth}{!}{%
\begin{tabular}{ccccc} \toprule
 &
  \multicolumn{2}{c}{\begin{tabular}[c]{@{}c@{}}ConditionA\\ Features\end{tabular}} &
  \multicolumn{2}{c}{\begin{tabular}[c]{@{}c@{}}ConditionB\\ Examples + Features\end{tabular}} \\ \midrule
 &
  Human &
  Human + AI &
  Human &
  Human + AI \\ \midrule
All &
  81 &
  \begin{tabular}[c]{@{}c@{}}78\\ (-3)\end{tabular} &
  86 &
  \begin{tabular}[c]{@{}c@{}}84\\ (-2)\end{tabular} \\ \midrule
\begin{tabular}[c]{@{}c@{}}Right\\ AIOutputs\end{tabular} &
  84 &
  \begin{tabular}[c]{@{}c@{}}88\\ (+4)\end{tabular} &
  87 &
  \begin{tabular}[c]{@{}c@{}}88\\ (+1)\end{tabular} \\ \midrule
\begin{tabular}[c]{@{}c@{}}Wrong\\ AIOutputs\end{tabular} &
  79 &
  \begin{tabular}[c]{@{}c@{}}68\\ (-11)\end{tabular} &
  86 &
  \begin{tabular}[c]{@{}c@{}}79\\ (-7)\end{tabular} \\ \bottomrule
\end{tabular}%
}
\end{table}

\textbf{Participants using `Examples + Features' had lower overreliance on AI} with lower performance degradation than those using only `Features'. Using all cases, participants had lower performance degradation (2\%) using `Examples + Features' than 'Features' (3\%). When `right' AI outputs were presented, participants' Human + AI performances were increased: 1\% performance increment using `Examples + Features' and 4\% performance increment using `Features'. When `wrong' AI outputs were presented, participants' Human + AI performances were decreased. Specifically, the participants using `Examples + Features' (7\%, $p<0.1$) had lower performance decrement than participants using `Features' (11\%, $p<0.05$).

\subsection{`Right' and `Wrong' Decisions}
Table \ref{tab:results_rightwrong} summarizes the ratios of `right' and `wrong' decisions on rehabilitation assessment tasks with AI outputs (`Human + AI') by participants. In addition, Table \ref{tab:results_rightwrong} describes the detailed ratios of `right' decisions (i.e. agreeing with `right' AI outputs and rejecting `wrong' AI outputs) and `wrong' decisions (i.e. agreeing with `wrong' AI outputs and rejecting `right' AI outputs).

\begin{table}[h]
\centering
\caption{The ratios and detailed analysis of `right' and `wrong' decisions by participants using AI with only Features (Condition A) and AI with Examples + Features (Condition B)}
\label{tab:results_rightwrong}
\resizebox{0.9\columnwidth}{!}{%
\begin{tabular}{ccc} \toprule
\textbf{} & \textbf{\begin{tabular}[c]{@{}c@{}}Condition A\\ Features\end{tabular}} & \textbf{\begin{tabular}[c]{@{}c@{}}ConditionB\\ Examples + Features\end{tabular}} \\ \midrule
\begin{tabular}[c]{@{}c@{}}Right\\ Decision\end{tabular}        & 88.9 & \begin{tabular}[c]{@{}c@{}}89.6\\ (+0.7)\end{tabular} \\ \midrule
\begin{tabular}[c]{@{}c@{}}Wrong\\ Decision\end{tabular}        & 11.1 & \begin{tabular}[c]{@{}c@{}}10.4\\ (-0.7)\end{tabular} \\ \midrule
\begin{tabular}[c]{@{}c@{}}Agree\\ RightAIOutputs\end{tabular} & 44.4 & \begin{tabular}[c]{@{}c@{}}44.4\\ \end{tabular}  \\ \midrule
\begin{tabular}[c]{@{}c@{}}Reject\\ WrongAIOutputs\end{tabular} & 44.4 & \begin{tabular}[c]{@{}c@{}}45.1\\ (0.7)\end{tabular}  \\ \midrule
\begin{tabular}[c]{@{}c@{}}Agree\\ WrongAIOutputs\end{tabular}  & 5.6  & \begin{tabular}[c]{@{}c@{}}4.9\\ (-0.7)\end{tabular}  \\ \midrule
\begin{tabular}[c]{@{}c@{}}Reject\\ RightAIOutputs\end{tabular} & 5.6 & \begin{tabular}[c]{@{}c@{}}5.6\\ \end{tabular}  \\ \bottomrule
\end{tabular}%
}
\end{table}

\textbf{Participants using `Examples + Features' had  0.7\% higher ratio of `right' decisions and 0.7\% lower ratio of `wrong' decisions than using `Features'}. For `right' decisions, participants using `Features' and `Examples + Features' had the same ratio of agreeing with `right' AI outputs and participants using \textbf{`Examples + Features' had a 0.7\% higher ratio of rejecting `wrong' AI outputs} than using `Features'. For `wrong' decisions, participants using \textbf{`Examples + Features' had a 0.7\% lower ratio of agreeing with `wrong' AI outputs} than using `Features' and the same ratio of rejecting `right' AI outputs using `Features'

\subsection{Duration}
Using the system with `Features', the participants spent an average of  60 seconds on an assessment. Using the system with `Examples + Features', the participants spent an average of 45 seconds. Overall, the usage of the system with only `Features' requires an average of 15 seconds more than that of the system with `Examples + Features' for a decision-making task.

\subsection{Discussion}
Our experimental results suggested that interactive example-based explanations for user's onboarding with AI (`Example + Features') are more effective for the domain experts, health professionals (i.e. therapists) to have a better-calibrated reliance on AI than presenting only an important feature explanation (`Features'). 
When the AI system presented `right' AI outputs to participants, the performance of human + AI has been improved compared to that of humans alone. When the AI system presented `wrong' AI outputs to participants, the performance of human + AI decreased compared to that of humans alone. Our study results follow the findings of the previous research on understanding the effect of AI explanations during the decision-support phase \cite{bansal2021does,lee2023counter}.

Compared to using only `Features', participants using `Examples + Features' had 2.7\% higher ratio of making `right' decisions, lower performance degradation and less overreliance on AI (Table \ref{tab:results_performance}). In addition, the participants using `Examples + Features' spent an average of 15 seconds less on an assessment task than those using only 'Features'. Our study results imply the potential of improving users' onboarding with interactive example-based explanations and their effective AI-assisted decision making than just using feature-based explanations during the decision-support phase. Our findings contrast with the findings of the previous study that during the decision-support phase, an example-based explanation underperformed a feature-based explanation to support calibrated trust and reliance on on AI \cite{wang2021explanations}.

Even if our results showed that interactive example-based explanations can be effective for therapists to improve their onboarding with AI and have a better-calibrated reliance on AI during human-AI collaborative decision-making, some participants were confused about how interactive parts of the system work. 
Thus, it is important to further explore how we can better educate and onboard with AI and AI explanations \cite{long2020ai}. 
Also, this work is limited to exploring the effect of interactive example-based explanations for user's onboarding with AI and does not provide generalization as we had a small size of participants and specified a particular AI/ML model (i.e. a feed-forward neural network model), the format of input data (i.e. videos), and a single clinical decision-making task (i.e. rehabilitation assessment). Thus, further studies are required to explore how to improve users' onboarding with AI for effective human-AI collaborative decision-making.

\section{Conclusion}
In this work, we contributed to an empirical study with domain experts, health professionals (i.e. therapists) to understand the effect of interactive example-based explanations to onboard users with AI for human-AI collaborative decision making. 
Our results showed that our proposed interactive example-based explanations (``Examples + Features'') during an onboarding phase assisted therapists to have a better-calibrated reliance on AI and to have a higher ratio of `right' decisions and a lower ratio of `wrong' decision 
than ``Features'' that presents only feature-based explanation during decision-support phase. 
We discuss the potential of using interactive AI explanations to support users' onboarding with AI for better-calibrated reliance on AI and challenges to improve human-AI collaborative decision-making.

%%%%%%%%%%%%%%%%%%%%%%%%%%%%%%%%%%%%%%%%%%%%%%%%%%%%%%%%%%%%%%%%%%%%%%%%

\begin{comment}
\begin{proof}
A full proof can be found in the supplementary material.
\end{proof}

Table captions should be centred \emph{above} the table, while figure 
captions should be centred \emph{below} the figure.\footnote{Footnotes
should be placed \emph{after} punctuation marks (such as full stops).}
 
\begin{table}[h]
\caption{Locations of selected conference editions.}
\centering
\begin{tabular}{ll@{\hspace{8mm}}ll} 
\toprule
AISB-1980 & Amsterdam & ECAI-1990 & Stockholm \\
ECAI-2000 & Berlin & ECAI-2010 & Lisbon \\
ECAI-2020 & \multicolumn{3}{l}{Santiago de Compostela (online)} \\
\bottomrule
\end{tabular}
\end{table}
\end{comment}
%%%%%%%%%%%%%%%%%%%%%%%%%%%%%%%%%%%%%%%%%%%%%%%%%%%%%%%%%%%%%%%%%%%%%%%%

%%%%%%%%%%%%%%%%%%%%%%%%%%%%%%%%%%%%%%%%%%%%%%%%%%%%%%%%%%%%%%%%%%%%%%%%

%%% Use this environment to include acknowledgements (optional).
%%% This will be omitted in doubleblind mode.

\begin{ack}
The authors thank all the participants in this work for their time and valuable inputs. This work is supported by the Singapore Ministry of Education (MOE) Academic Research Fund (AcRF) Tier 1 grant.
\end{ack}

%%%%%%%%%%%%%%%%%%%%%%%%%%%%%%%%%%%%%%%%%%%%%%%%%%%%%%%%%%%%%%%%%%%%%%%%
\begin{table}[htp]
\centering
\caption{Detailed Demographics of Therapists who have experience in stroke rehabilitation (P1 - P10) for the Interviews.}
\label{tab:demographics_detailed_interview}
\resizebox{1.0\columnwidth}{!}{%
\begin{tabular}{lllllll} \toprule
PID &
  Sex &
  Age &
  Occuptation &
  Setting &
  \# of yrs \\ \midrule
P1 &
  Female &
  25 - 34 years &
  PhysioTherapist (PT) &
  Outpatient Clinic &
  7  \\ \midrule
P2 &
  Male &
  25 - 34 years &
  PhysioTherapist (PT) &
  Inpatient Rehabilitation &
  2 \\ \midrule
P3 &
  Male &
  25 - 34 years &
  PhysioTherapist (PT) &
  Home Care &
  8 \\ \midrule
P4 &
  Female &
  35 - 44 years &
  PhysioTherapist (PT) &
  Outpatient Clinic &
  11  \\ \midrule
P5 &
  Female &
  25 - 34 years &
  PhysioTherapist (PT) &
  Inpatient Rehabilitation &
  9 \\ \midrule
P6 &
  Female &
  45 - 54 years &
  PhysioTherapist (PT) &
  Skilled Nursing Facility &
  30  \\ \midrule
P7 &
  Female &
  35 - 44 years &
  Occupational Therapist (OT) &
  Outpatient Clinic &
  14 \\ \midrule
P8 &
  Female &
  35 - 44 years &
  Occupational Therapist (OT) &
  Homecare &
  11  \\ \midrule
P9 &
  Female &
  25 - 34 years &
  Occupational Therapist (OT) &
  Skilled Nursing Facility &
  6 \\ \midrule
P10 &
  Female &
  25 - 34 years &
  Occupational Therapist (OT) &
  Inpatient Rehabilitation &
  5 \\ \bottomrule
\end{tabular}%
}
\end{table}

% \usepackage{graphicx}
\begin{table}[htp]
\centering
\caption{Detailed Demographics of Therapists who have experience in stroke rehabilitation (TP1 - TP16) for the User Study.}
\label{tab:demographic_detailed}
\resizebox{1.0\columnwidth}{!}{%
\begin{tabular}{cccllc} \toprule
PID &
  Sex &
  Age &
  \multicolumn{1}{c}{Occupation} &
  Setting &
  \# of yrs \\ \midrule
TP1  & Female & 18 - 24 years & Occupational Therapist       & Acute \& Inpatient        & <1      \\
TP2  & Female & 25 - 34 years & Occupational Therapist       & Inpatient Rehabilitation & 5       \\
TP3  & Male   & 25 - 34 years & Physiotherapist              & Acute Care               & 0.25    \\
TP4  & Female & 35 - 44 years & Physiotherapist              & Outpatient Clinic        & 11      \\
TP5  & Female & 25 - 34 years & Occupational Therapist       & Inpatient Rehabilitation & 6       \\
TP6  & Female & 25 - 34 years & Occupational Therapist       & Skilled Nursing Facility & 4       \\
TP7  & Female & 25 - 34 years & Occupational Therapist       & Outpatient Clinic        & 5       \\
TP8  & Female & 25 - 34 years & Occupational Therapist       & Inpatient Rehabilitation & 9       \\
TP9  & Male   & 25 - 34 years & Occupational Therapist       & Skilled Nursing Facility & 5       \\
TP10 & Female & 25 - 34 years & Occupational Therapist       & Inpatient Rehabilitation & 10      \\
TP11 & Female & 25 - 34 years & Occupational Therapist       & Inpatient Rehabilitation & 4       \\
TP12 & Male   & 25 - 34 years & Physiotherapist              & Inpatient Rehabilitation & 1.3     \\
TP13 & Female & 35 - 44 years & Occupational Therapist       & Outpatient Clinic        & 14      \\
TP14 & Female & 35 - 44 years & Physiotherapist              & Skilled Nursing Facility & 2       \\
TP15 & Female & 25 - 34 years & Physiotherapist              & Home Care                & 12      \\
TP16 & Female & 45 - 54 years & Physiotherapist              & Skilled Nursing Facility & 30      \\
%TP17 & Female & 25 - 34 years & Physiotherapist              & Inpatient Rehabilitation & 9       \\ 
 %\\ 
 \bottomrule
\end{tabular}% 
}
\end{table}

\newpage
\bibliography{mybibfile}

\begin{thebibliography}{46}
\providecommand{\natexlab}[1]{#1}
\providecommand{\url}[1]{\texttt{#1}}
\expandafter\ifx\csname urlstyle\endcsname\relax
  \providecommand{\doi}[1]{doi: #1}\else
  \providecommand{\doi}{doi: \begingroup \urlstyle{rm}\Url}\fi

\bibitem[Abdi and Williams(2010)]{abdi2010principal}
H.~Abdi and L.~J. Williams.
\newblock Principal component analysis.
\newblock \emph{Wiley interdisciplinary reviews: computational statistics}, 2\penalty0 (4):\penalty0 433--459, 2010.

\bibitem[Amershi et~al.(2019)Amershi, Weld, Vorvoreanu, Fourney, Nushi, Collisson, Suh, Iqbal, Bennett, Inkpen, et~al.]{amershi2019guidelines}
S.~Amershi, D.~Weld, M.~Vorvoreanu, A.~Fourney, B.~Nushi, P.~Collisson, J.~Suh, S.~Iqbal, P.~N. Bennett, K.~Inkpen, et~al.
\newblock Guidelines for human-ai interaction.
\newblock In \emph{Proceedings of the 2019 chi conference on human factors in computing systems}, pages 1--13, 2019.

\bibitem[Bansal et~al.(2021)Bansal, Wu, Zhou, Fok, Nushi, Kamar, Ribeiro, and Weld]{bansal2021does}
G.~Bansal, T.~Wu, J.~Zhou, R.~Fok, B.~Nushi, E.~Kamar, M.~T. Ribeiro, and D.~Weld.
\newblock Does the whole exceed its parts? the effect of ai explanations on complementary team performance.
\newblock In \emph{Proceedings of the 2021 CHI Conference on Human Factors in Computing Systems}, pages 1--16, 2021.

\bibitem[Becht et~al.(2019)Becht, McInnes, Healy, Dutertre, Kwok, Ng, Ginhoux, and Newell]{becht2019dimensionality}
E.~Becht, L.~McInnes, J.~Healy, C.-A. Dutertre, I.~W. Kwok, L.~G. Ng, F.~Ginhoux, and E.~W. Newell.
\newblock Dimensionality reduction for visualizing single-cell data using umap.
\newblock \emph{Nature biotechnology}, 37\penalty0 (1):\penalty0 38--44, 2019.

\bibitem[Beede et~al.(2020)Beede, Baylor, Hersch, Iurchenko, Wilcox, Ruamviboonsuk, and Vardoulakis]{beede2020human}
E.~Beede, E.~Baylor, F.~Hersch, A.~Iurchenko, L.~Wilcox, P.~Ruamviboonsuk, and L.~M. Vardoulakis.
\newblock A human-centered evaluation of a deep learning system deployed in clinics for the detection of diabetic retinopathy.
\newblock In \emph{Proceedings of the 2020 CHI conference on human factors in computing systems}, pages 1--12, 2020.

\bibitem[Boggust et~al.(2022)Boggust, Carter, and Satyanarayan]{boggust2022embedding}
A.~Boggust, B.~Carter, and A.~Satyanarayan.
\newblock Embedding comparator: Visualizing differences in global structure and local neighborhoods via small multiples.
\newblock In \emph{27th international conference on intelligent user interfaces}, pages 746--766, 2022.

\bibitem[Bu{\c{c}}inca et~al.(2021)Bu{\c{c}}inca, Malaya, and Gajos]{buccinca2021trust}
Z.~Bu{\c{c}}inca, M.~B. Malaya, and K.~Z. Gajos.
\newblock To trust or to think: cognitive forcing functions can reduce overreliance on ai in ai-assisted decision-making.
\newblock \emph{Proceedings of the ACM on Human-Computer Interaction}, 5\penalty0 (CSCW1):\penalty0 1--21, 2021.

\bibitem[Bussone et~al.(2015)Bussone, Stumpf, and O'Sullivan]{bussone2015role}
A.~Bussone, S.~Stumpf, and D.~O'Sullivan.
\newblock The role of explanations on trust and reliance in clinical decision support systems.
\newblock In \emph{2015 international conference on healthcare informatics}, pages 160--169. IEEE, 2015.

\bibitem[Cai et~al.(2019{\natexlab{a}})Cai, Jongejan, and Holbrook]{cai2019effects}
C.~J. Cai, J.~Jongejan, and J.~Holbrook.
\newblock The effects of example-based explanations in a machine learning interface.
\newblock In \emph{Proceedings of the 24th international conference on intelligent user interfaces}, pages 258--262, 2019{\natexlab{a}}.

\bibitem[Cai et~al.(2019{\natexlab{b}})Cai, Reif, Hegde, Hipp, Kim, Smilkov, Wattenberg, Viegas, Corrado, Stumpe, et~al.]{cai2019human}
C.~J. Cai, E.~Reif, N.~Hegde, J.~Hipp, B.~Kim, D.~Smilkov, M.~Wattenberg, F.~Viegas, G.~S. Corrado, M.~C. Stumpe, et~al.
\newblock Human-centered tools for coping with imperfect algorithms during medical decision-making.
\newblock In \emph{Proceedings of the 2019 chi conference on human factors in computing systems}, pages 1--14, 2019{\natexlab{b}}.

\bibitem[Cai et~al.(2019{\natexlab{c}})Cai, Winter, Steiner, Wilcox, and Terry]{cai2019hello}
C.~J. Cai, S.~Winter, D.~Steiner, L.~Wilcox, and M.~Terry.
\newblock " hello ai": uncovering the onboarding needs of medical practitioners for human-ai collaborative decision-making.
\newblock \emph{Proceedings of the ACM on Human-computer Interaction}, 3\penalty0 (CSCW):\penalty0 1--24, 2019{\natexlab{c}}.

\bibitem[Gilpin et~al.(2018)Gilpin, Bau, Yuan, Bajwa, Specter, and Kagal]{gilpin2018explaining}
L.~H. Gilpin, D.~Bau, B.~Z. Yuan, A.~Bajwa, M.~Specter, and L.~Kagal.
\newblock Explaining explanations: An overview of interpretability of machine learning.
\newblock In \emph{2018 IEEE 5th International Conference on data science and advanced analytics (DSAA)}, pages 80--89. IEEE, 2018.

\bibitem[Gladstone et~al.(2002)Gladstone, Danells, and Black]{gladstone2002fugl}
D.~J. Gladstone, C.~J. Danells, and S.~E. Black.
\newblock The fugl-meyer assessment of motor recovery after stroke: a critical review of its measurement properties.
\newblock \emph{Neurorehabilitation and neural repair}, 16\penalty0 (3):\penalty0 232--240, 2002.

\bibitem[Grinberg(2018)]{grinberg2018flask}
M.~Grinberg.
\newblock \emph{Flask web development: developing web applications with python}.
\newblock " O'Reilly Media, Inc.", 2018.

\bibitem[Holstein et~al.(2023)Holstein, De-Arteaga, Tumati, and Cheng]{holstein2023toward}
K.~Holstein, M.~De-Arteaga, L.~Tumati, and Y.~Cheng.
\newblock Toward supporting perceptual complementarity in human-ai collaboration via reflection on unobservables.
\newblock \emph{Proceedings of the ACM on Human-Computer Interaction}, 7\penalty0 (CSCW1):\penalty0 1--20, 2023.

\bibitem[Khairat et~al.(2018)Khairat, Marc, Crosby, Al~Sanousi, et~al.]{khairat2018reasons}
S.~Khairat, D.~Marc, W.~Crosby, A.~Al~Sanousi, et~al.
\newblock Reasons for physicians not adopting clinical decision support systems: critical analysis.
\newblock \emph{JMIR medical informatics}, 6\penalty0 (2):\penalty0 e8912, 2018.

\bibitem[Kohli and Jha(2018)]{kohli2018cad}
A.~Kohli and S.~Jha.
\newblock Why cad failed in mammography.
\newblock \emph{Journal of the American College of Radiology}, 15\penalty0 (3):\penalty0 535--537, 2018.

\bibitem[Lai and Tan(2019)]{lai2019human}
V.~Lai and C.~Tan.
\newblock On human predictions with explanations and predictions of machine learning models: A case study on deception detection.
\newblock In \emph{Proceedings of the conference on fairness, accountability, and transparency}, pages 29--38, 2019.

\bibitem[Lai et~al.(2021)Lai, Chen, Liao, Smith-Renner, and Tan]{lai2021towards}
V.~Lai, C.~Chen, Q.~V. Liao, A.~Smith-Renner, and C.~Tan.
\newblock Towards a science of human-ai decision making: a survey of empirical studies.
\newblock \emph{arXiv preprint arXiv:2112.11471}, 2021.

\bibitem[Lakkaraju et~al.(2020)Lakkaraju, Adebayo, and Singh]{lakkaraju2020explaining}
H.~Lakkaraju, J.~Adebayo, and S.~Singh.
\newblock Explaining machine learning predictions: State-of-the-art, challenges, and opportunities.
\newblock \emph{NeurIPS Tutorial}, 2020.

\bibitem[Lee(2024)]{lee2024towards}
M.~H. Lee.
\newblock Towards gradient-based time-series explanations through a spatiotemporal attention network.
\newblock \emph{arXiv preprint arXiv:2405.17444}, 2024.

\bibitem[Lee and Chew(2023)]{lee2023counter}
M.~H. Lee and C.~J. Chew.
\newblock Understanding the effect of counterfactual explanations on trust and reliance on ai for human-ai collaborative clinical decision making.
\newblock \emph{Proc. ACM Hum.-Comput. Interact.}, 7\penalty0 (CSCW2), oct 2023.

\bibitem[Lee et~al.(2019)Lee, Siewiorek, Smailagic, Bernardino, and Badia]{lee2019learning}
M.~H. Lee, D.~P. Siewiorek, A.~Smailagic, A.~Bernardino, and S.~B.~i. Badia.
\newblock Learning to assess the quality of stroke rehabilitation exercises.
\newblock In \emph{Proceedings of the 24th International Conference on Intelligent User Interfaces}, pages 218--228, 2019.

\bibitem[Lee et~al.(2020)Lee, Siewiorek, Smailagic, Bernardino, and Berm{\'u}dez~i Badia]{lee2020co}
M.~H. Lee, D.~P. Siewiorek, A.~Smailagic, A.~Bernardino, and S.~Berm{\'u}dez~i Badia.
\newblock Co-design and evaluation of an intelligent decision support system for stroke rehabilitation assessment.
\newblock \emph{Proceedings of the ACM on Human-Computer Interaction}, 4\penalty0 (CSCW2):\penalty0 1--27, 2020.

\bibitem[Lee et~al.(2021)Lee, Siewiorek, Smailagic, Bernardino, and Berm{\'u}dez~i Badia]{lee2021human}
M.~H. Lee, D.~P. Siewiorek, A.~Smailagic, A.~Bernardino, and S.~Berm{\'u}dez~i Badia.
\newblock A human-ai collaborative approach for clinical decision making on rehabilitation assessment.
\newblock In \emph{Proceedings of the 2021 CHI conference on human factors in computing systems}, pages 1--14, 2021.

\bibitem[Lee et~al.(2024)Lee, Choo, and Thilarajah]{lee2024improving}
M.~H. Lee, S.~X.~Y. Choo, and S.~D. Thilarajah.
\newblock Improving health professionals' onboarding with ai and xai for trustworthy human-ai collaborative decision making.
\newblock \emph{arXiv preprint arXiv:2405.16424}, 2024.

\bibitem[Liao and Varshney(2021)]{liao2021human}
Q.~V. Liao and K.~R. Varshney.
\newblock Human-centered explainable ai (xai): From algorithms to user experiences.
\newblock \emph{arXiv preprint arXiv:2110.10790}, 2021.

\bibitem[Lipton(2018)]{lipton2018mythos}
Z.~C. Lipton.
\newblock The mythos of model interpretability: In machine learning, the concept of interpretability is both important and slippery.
\newblock \emph{Queue}, 16\penalty0 (3):\penalty0 31--57, 2018.

\bibitem[Liu et~al.(2016)Liu, Maljovec, Wang, Bremer, and Pascucci]{liu2016visualizing}
S.~Liu, D.~Maljovec, B.~Wang, P.-T. Bremer, and V.~Pascucci.
\newblock Visualizing high-dimensional data: Advances in the past decade.
\newblock \emph{IEEE transactions on visualization and computer graphics}, 23\penalty0 (3):\penalty0 1249--1268, 2016.

\bibitem[London(2019)]{london2019artificial}
A.~J. London.
\newblock Artificial intelligence and black-box medical decisions: accuracy versus explainability.
\newblock \emph{Hastings Center Report}, 49\penalty0 (1):\penalty0 15--21, 2019.

\bibitem[Long and Magerko(2020)]{long2020ai}
D.~Long and B.~Magerko.
\newblock What is ai literacy? competencies and design considerations.
\newblock In \emph{Proceedings of the 2020 CHI conference on human factors in computing systems}, pages 1--16, 2020.

\bibitem[Lundberg and Lee(2017)]{NIPS2017Shap}
S.~M. Lundberg and S.-I. Lee.
\newblock A unified approach to interpreting model predictions.
\newblock In I.~Guyon, U.~V. Luxburg, S.~Bengio, H.~Wallach, R.~Fergus, S.~Vishwanathan, and R.~Garnett, editors, \emph{Advances in Neural Information Processing Systems 30}, pages 4765--4774. Curran Associates, Inc., 2017.

\bibitem[Mitchell et~al.(2019)Mitchell, Wu, Zaldivar, Barnes, Vasserman, Hutchinson, Spitzer, Raji, and Gebru]{mitchell2019model}
M.~Mitchell, S.~Wu, A.~Zaldivar, P.~Barnes, L.~Vasserman, B.~Hutchinson, E.~Spitzer, I.~D. Raji, and T.~Gebru.
\newblock Model cards for model reporting.
\newblock In \emph{Proceedings of the conference on fairness, accountability, and transparency}, pages 220--229, 2019.

\bibitem[PAIR.(2019)]{GooglePAIR2019}
G.~PAIR.
\newblock "people + ai guidebook, 2019.
\newblock URL \url{https://pair.withgoogle.com/guidebook/}.

\bibitem[Paszke et~al.(2019)Paszke, Gross, Massa, Lerer, Bradbury, Chanan, Killeen, Lin, Gimelshein, Antiga, et~al.]{paszke2019pytorch}
A.~Paszke, S.~Gross, F.~Massa, A.~Lerer, J.~Bradbury, G.~Chanan, T.~Killeen, Z.~Lin, N.~Gimelshein, L.~Antiga, et~al.
\newblock Pytorch: An imperative style, high-performance deep learning library.
\newblock \emph{Advances in neural information processing systems}, 32, 2019.

\bibitem[Preece(2018)]{preece2018asking}
A.~Preece.
\newblock Asking ‘why’in ai: Explainability of intelligent systems--perspectives and challenges.
\newblock \emph{Intelligent Systems in Accounting, Finance and Management}, 25\penalty0 (2):\penalty0 63--72, 2018.

\bibitem[Rajpurkar et~al.(2022)Rajpurkar, Chen, Banerjee, and Topol]{rajpurkar2022ai}
P.~Rajpurkar, E.~Chen, O.~Banerjee, and E.~J. Topol.
\newblock Ai in health and medicine.
\newblock \emph{Nature medicine}, 28\penalty0 (1):\penalty0 31--38, 2022.

\bibitem[Ribeiro et~al.(2016)Ribeiro, Singh, and Guestrin]{ribeiro2016should}
M.~T. Ribeiro, S.~Singh, and C.~Guestrin.
\newblock " why should i trust you?" explaining the predictions of any classifier.
\newblock In \emph{Proceedings of the 22nd ACM SIGKDD international conference on knowledge discovery and data mining}, pages 1135--1144, 2016.

\bibitem[Sutton et~al.(2020)Sutton, Pincock, Baumgart, Sadowski, Fedorak, and Kroeker]{sutton2020overview}
R.~T. Sutton, D.~Pincock, D.~C. Baumgart, D.~C. Sadowski, R.~N. Fedorak, and K.~I. Kroeker.
\newblock An overview of clinical decision support systems: benefits, risks, and strategies for success.
\newblock \emph{NPJ digital medicine}, 3\penalty0 (1):\penalty0 17, 2020.

\bibitem[Topol(2019)]{topol2019high}
E.~J. Topol.
\newblock High-performance medicine: the convergence of human and artificial intelligence.
\newblock \emph{Nature medicine}, 25\penalty0 (1):\penalty0 44--56, 2019.

\bibitem[Turney and Pantel(2010)]{turney2010frequency}
P.~D. Turney and P.~Pantel.
\newblock From frequency to meaning: Vector space models of semantics.
\newblock \emph{Journal of artificial intelligence research}, 37:\penalty0 141--188, 2010.

\bibitem[Van~der Maaten and Hinton(2008)]{van2008visualizing}
L.~Van~der Maaten and G.~Hinton.
\newblock Visualizing data using t-sne.
\newblock \emph{Journal of machine learning research}, 9\penalty0 (11), 2008.

\bibitem[Vasconcelos et~al.(2023)Vasconcelos, J{\"o}rke, Grunde-McLaughlin, Gerstenberg, Bernstein, and Krishna]{vasconcelos2023explanations}
H.~Vasconcelos, M.~J{\"o}rke, M.~Grunde-McLaughlin, T.~Gerstenberg, M.~S. Bernstein, and R.~Krishna.
\newblock Explanations can reduce overreliance on ai systems during decision-making.
\newblock \emph{Proceedings of the ACM on Human-Computer Interaction}, 7\penalty0 (CSCW1):\penalty0 1--38, 2023.

\bibitem[Wang et~al.(2019)Wang, Yang, Abdul, and Lim]{wang2019designing}
D.~Wang, Q.~Yang, A.~Abdul, and B.~Y. Lim.
\newblock Designing theory-driven user-centric explainable ai.
\newblock In \emph{Proceedings of the 2019 CHI conference on human factors in computing systems}, pages 1--15, 2019.

\bibitem[Wang and Yin(2021)]{wang2021explanations}
X.~Wang and M.~Yin.
\newblock Are explanations helpful? a comparative study of the effects of explanations in ai-assisted decision-making.
\newblock In \emph{26th international conference on intelligent user interfaces}, pages 318--328, 2021.

\bibitem[Zavr{\v{s}}nik(2020)]{zavrvsnik2020criminal}
A.~Zavr{\v{s}}nik.
\newblock Criminal justice, artificial intelligence systems, and human rights.
\newblock In \emph{ERA forum}, volume~20, pages 567--583. Springer, 2020.

\end{thebibliography}

\end{document}